\newcommand\SCO{SrCuO$_2$\xspace}
\newcommand\SCON{SrCu$_{0.99}$Ni$_{0.01}$O$_2$\xspace}
\newcommand\SCONX{SrCu$_{1-x}$Ni$_{x}$O$_2$\xspace}
\newcommand\SCOC{Sr$_{1-x}$Ca$_x$CuO$_2$\xspace}

\documentclass[prl,twocolumn,floatfix,preprintnumbers,amsmath,amssymb,superscriptaddress]{revtex4}

\usepackage{graphicx}
\usepackage{dcolumn}
\usepackage{bm}
\usepackage{color}
\usepackage{xspace}

\begin{document}

\title{Spin pseudogap in Ni-doped SrCuO$_2$}

\author{G. Simutis}
\affiliation{Neutron Scattering and Magnetism, Laboratory for Solid State
Physics, ETH Z\"urich, Z\"urich, Switzerland}

\author{S. Gvasaliya}
\affiliation{Neutron Scattering and Magnetism, Laboratory for Solid State
Physics, ETH Z\"urich, Z\"urich, Switzerland}

\author{M. M{\aa}nsson}
\altaffiliation[Present address: ]{Laboratory for Quantum Magnetism (LQM)
\'Ecole polytechnique f\'ed\'erale de Lausanne (EPFL), Lausanne, Switzerland and Laboratory for Neutron Scattering, Paul Scherrer Insitut, Villigen, PSI, Switzerland}

\affiliation{Neutron Scattering and Magnetism, Laboratory for Solid State
Physics, ETH Z\"urich, Z\"urich, Switzerland}

\author{A. L. Chernyshev}
\affiliation{Department of Physics and Astronomy, University of California, Irvine, California 92697, USA}

\author{A. Mohan}
\affiliation{Leibniz Institute for Solid State and Materials Research IFW Dresden, P.O. Box 270116, D-01171 Dresden, Germany}

\author{S. Singh}
\affiliation{Leibniz Institute for Solid State and Materials Research IFW Dresden, P.O. Box 270116, D-01171 Dresden, Germany}

\author{C. Hess}
\affiliation{Leibniz Institute for Solid State and Materials Research IFW Dresden, P.O. Box 270116, D-01171 Dresden, Germany}

\author{A. T. Savici}
\affiliation{NScD, Oak Ridge National
Laboratory, Oak Ridge, TN 37831, USA}

\author{A. I. Kolesnikov}
\affiliation{NScD, Oak Ridge National
Laboratory, Oak Ridge, TN 37831, USA}

\author{A. Piovano}
\affiliation{Institut Laue-Langevin, 6 rue Jules Horowitz, 38042 Grenoble Cedex 9, France}

\author{T. Perring}
\affiliation{ISIS Facility, Rutherford Appleton Laboratory, Chilton, Didcot, Oxon OX11 OQX, United Kingdom}

\author{I. Zaliznyak}
\affiliation{ Brookhaven National Laboratory, Upton, New York 11973, USA}

\author{B. B\"uchner }
\affiliation{Leibniz Institute for Solid State and Materials Research IFW Dresden, P.O. Box 270116, D-01171 Dresden, Germany}
\affiliation{Department of Physics, TU Dresden, D-01069 Dresden, Germany}

\author{A. Zheludev}
 \email{zhelud@ethz.ch}
 \homepage{http://www.neutron.ethz.ch/}
\affiliation{Neutron Scattering and Magnetism, Laboratory for Solid State
Physics, ETH Z\"urich, Z\"urich, Switzerland}

\date{June 29, 2013}

\begin{abstract}
The $S=1/2$ spin chain material \SCO doped with 1\% $S=1$ Ni-impurities is studied by inelastic neutron scattering. At low temperatures, the spectrum shows a pseudogap $\Delta\approx 8$~meV, absent in the parent compound, and not related to any structural phase transition. The pseudogap is shown to be a generic feature of quantum spin chains with dilute defects. A simple model based on this idea quantitatively accounts for the exprimental data measured in the temperature range 2-300~K, and allows to represent the momentum-integrated dynamic structure factor in a universal scaling form.
\end{abstract}

 \maketitle

Due to the unique topology of one dimensional space, defects in one dimensional systems often have a profound effect on the physical properties. Indeed, regardless of their size, defects can not be avoided by propagating quasiparticles \cite{Kane1992}. In certain cases, disorder may qualitatively reconstruct the ground state\cite{Giamarchi1987,Fisher1994,Damle2000} as well as low energy excitations \cite{Schmitteckert1998,Motrunich2001}, and to drastically influence transport phenomena\cite{Motrunich2001,Gornyi2005,Karahalios2009}. Such behavior has been extensively studied in numerous systems, including one-dimensional quantum magnets. A particularly important case is that of a simple antiferromagnetic (AF) Heisenberg $S=1/2$ chain, which is a cornerstone example of quantum criticality with scale-free AF correlations \cite{Tsvelikbook,Giamarchibook}. Due to quantum fluctuations, the defect-free chain shows no magnetic ordering, and a finite magnetic susceptibility even at zero temperature. Defects will disrupt the local AF correlations, and thereby liberate additional low-energy spin degrees of freedom. This handwaving consideration can actually be made rigorous through a real space Renormalization Group (RG) argument \cite{Dasgupta1980}. Predictions of the corresponding strong-disorder RG theory include divergent low-energy density of states and spin structure factor \cite{Damle2000}. These conclusions are universal in that they are independent of the actual details of disorder. By now, several experimental confirmations of such random singlet(RS) states have been obtained in prototype spin chain materials with compositional randomness \cite{Tippie1981,Shiroka2011}.

In view of the above, it may appear surprising that NMR experiments on one of the best knows spin chain materials \SCO \cite{Rice1997,Motoyama1996,Zaliznyak1999,Zaliznyak2004} showed the opening of a spin gap in samples disordered by low level Ca-doping \cite{Hammerath2011}. Ca replaces Sr, and therefore does not affect the chains directly. A gap in the spin excitation spectrum implies a {\it depletion} of low-energy magnetic states. The unusual effect was tentatively attributed to the influence of Ca-doping on the crystalline lattice, leading to  a spin-Peierls phase transition of the kind seen in CuGeO$_3$ \cite{Uchinokura2002}. The focus of the present paper is \SCONX, a Ni-doped version of the same parent material. The $S=1$ Ni$^{2+}$ impurities are directly inserted into the spin chains, replacing the $S=1/2$ Cu$^{2+}$ ions.  We use neutron spectroscopy to show that Ni-doping depletes the low-energy spin excitation spectrum, opening a {\it spin-pseudogap}. We then argue that such a pseudogap is a {\it  generic} spectral feature of $S=1/2$ AF chains with a low concentration of weak links or defects, a situation which is far from the strong-disorder RG fixed point. Finally, we demonstrate that defects do not entirely wipe out
the universal scaling properties of the correlation functions.
The latter are only ``masked'' by a defect-specific (non-universal) envelope function
and can be recovered.

As discussed in \cite{Zaliznyak1999}, \SCO has a double $S=1/2$-chain structure \footnote{space group $Cmcm$, $a=3.556$~\AA, $b=16.27$~\AA and $c=3.904$~\AA}. The chains run along the $c$ axis of the orthorhombic crystal. By now it is well established that the in-chain AF coupling $J=226$~meV is dominant. The fully frustrated ferromagnetic inter-chain interaction is considerably weaker, of the order of $J'\sim 10$~meV \cite{Gopalan1994,Rice1997}. As a result, the spin excitation spectrum, as determined with great accuracy and quantitatively with neutron spectroscopy \cite{Zaliznyak1999,Zaliznyak2004}, is essentially that of isolated spin chains. Deviations emerge only at energies below $\sim 1$~meV and are due to three-dimensional interactions that lead to very weak magnetic ordering below $T_\mathrm{N}\sim 5$~K \cite{Zaliznyak1999}. In the present work we study the 1\% Ni-doped version of such parent compound, namely \SCON. Four single crystal samples of total mass 4~g and a mosaic spread of 0.9$^\circ$ were grown using the floating zone technique. All neutron measurements described below were performed with the sample mounted with the $b$ axis normal to the horizontal scattering plane of the spectrometer. Initial time of flight (TOF) thermal neutron data were taken on MAPS spectrometer at ISIS, though most measurements were performed using the SEQUOIA \cite{Granroth2010} instrument at ORNL (incident neutron energy $E_i=25$~meV). Thermal three-axis neutron experiments were carried out at the IN8 spectrometer at ILL (final neutron energy $E_f=14.7$~meV). Cold neutron data were acquired at the TASP 3-axis spectrometer of PSI ($E_f=5$~meV).

\begin{figure}
\includegraphics[width=\columnwidth]{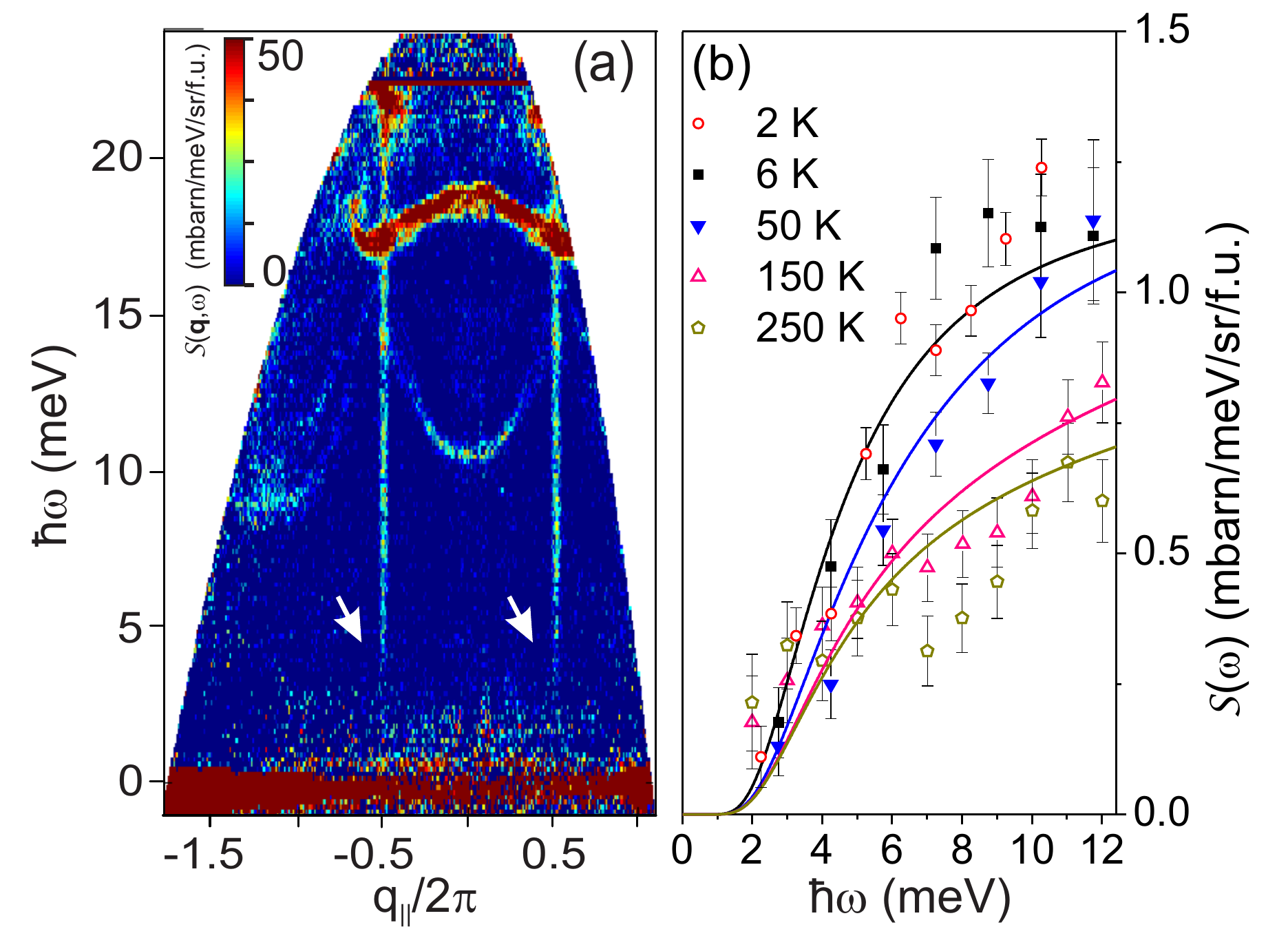}
\caption{(Color online) (a) Background-subtracted time-of-flight (TOF) neutron spectrum measured in \SCON on SEQUOIA, as a function of momentum transfer along the spin chains and energy transfer. Arrows indicate the spin pseudogap. (b) Momentum-integrated dynamic structure factor measured in \SCON at several temperatures, using TOF (solid symbols) and 3-axis (open symbols) spectroscopy. The lines are calculated using Eq.(~\ref{SW}), assuming $x=0.017$. \label{fig:data}}
\end{figure}

As an overview of the scattering, the TOF spectrum collected at $T=6$~K is shown in Fig.~\ref{fig:data}(a) in projection onto the plane of energy transfer $\hbar\omega$ and momentum transfer along the chain axis $q_\|\equiv \mathbf{qc}$. Here the background from the sample-can and instrument components has been measured separately and subtracted, and the signal has been normalized using a vanadium standard. What is plotted is therefore the dynamic structure factor $\mathcal{S}(\mathbf{q},\omega)$ of the sample, normalized to formula unit. The features that are most relevant to our problem are the two vertical streaks of intensity at $q_\|=\pm \pi$. This is the very steeply dispersive magnetic scattering near the AF zone-centers, as previously observed in the parent material \cite{Zaliznyak2004}. Above roughly 8~meV energy transfer some low-lying optical phonon branches start to overlap with the magnetic signal. Because of that, a quantitative determination of the magnetic component above 12~meV becomes problematic.

The main experimental result of this study is the measured momentum-integrated dynamic structure factor $\mathcal{S}(\omega)=\int \mathrm{d}q_\| \mathcal{S}(q_\|,\omega)$. The integral is taken in the vicinity of $q_\|=\pi$ , and quantifies one-dimensional magnetic correlations \cite{Dender1996,Dender1997thesis,Zheludev2007-2}. Experimentally, it was obtained from the TOF data by fitting Gaussian profiles to constant-energy cuts across the measured spectrum, and taking the peak area vs. energy transfer. The approach is justified, as no 3-dimensional correlations are expected  above $\approx 1$~meV \cite{Zaliznyak1999}. The result  is shown for several temperatures in Fig.~\ref{fig:data}(b). Here we also show data sets obtained using the 3-axis spectrometers. Due to the very steep dispersion and a rather coarse experimental wave vector resolution, 3-axis constant-$q$ scans at $q_\|=\pm \pi$  are automatically $q_\|$-integrated. The background for point-by-point subtraction was measured off the AF zone-center, at $q_\|=\pm 0.8\pi$ and $q_\|=\pm 1.2\pi$. All scans measured with the 3-axis technique were scaled to match the normalization of the TOF data, using temperatures where spectra were measured using both methods. In all cases, care was taken to correct the measured intensities for the magnetic form factor of Cu$^{2+}$ \footnote{Note that for the TOF data this correction is not entirely trivial, as the form factor depends on the absolute wave vector transfer, and has to be applied before the the $q_\|$ projection.}.

\begin{figure}
\includegraphics[width=\columnwidth]{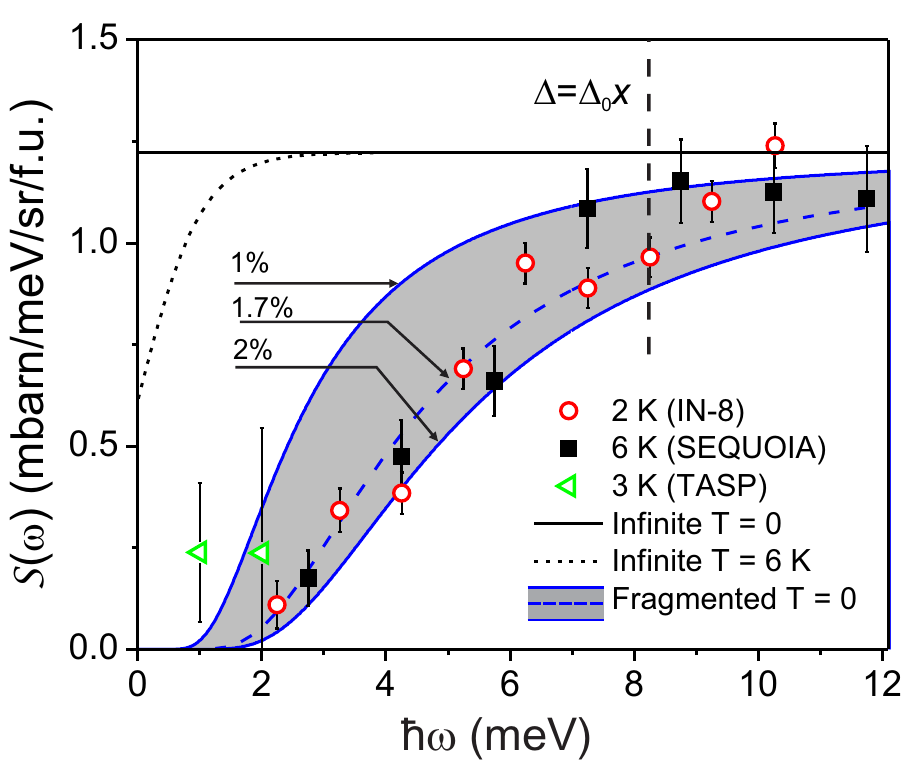}
\caption{(Color online) Symbols: energy dependence  of the momentum-integrated dynamic structure factor measured in \SCON at $T=2$~K and $T=6$~K. The thin solid and dotted lines are the theoretical expectation for a defect-free spin chain at $T=0$ and $T=6$~K, respectively. The solid and dashed curves are theoretical predictions for doped chains, assuming $x=0.01$, $x=0.017$ and $x=0.02$, as discussed in the text.\label{fig:models}}
\end{figure}

In Fig.~\ref{fig:models} we plot $\mathcal{S}(\omega)$ measured in \SCON below $T=6$~K (symbols), in comparison with the theoretical expectation, normalized per Cu$^{2+}$ ion,
 \begin{equation}
\mathcal{S}_\infty(\omega)=(\gamma r_0)^2\frac{k_f}{k_i}\frac{2g^2}{4}A\frac{n(\omega)+1}{\pi J}\tanh\left(\frac{\hbar \omega}{2\kappa_\mathrm{B} T}\right)\label{sqw}
\end{equation}
for an infinite-length defect-free $S=1/2$ Heisenberg AF spin  chain \cite{Muller1981,Schulz1986,Dender1997,Starykh1998,Zaliznyak2004} at $T=0$ (solid line) and $T=6$~K (dotted line). In Eq.~(\ref{sqw}), $(\gamma r_0)^2=0.290$~barn, $k_i$ and $k_f$ are the incident and final neutron wave vectors, and $A=1.34$ is the M\"uller ansatz normalization from \cite{Muller1981}. In all our calculations we have assumed $g^2=(2.12)^2$, which is the average value for the spin components probed in our experiments, as calculated based on the results of ESR measurements \cite{Kataev2001}. Note that the theoretical curves are plotted without any arbitrary scaling factor, which is possible thanks to experimental data also being normalized to unit formula. Compared to a defect-free chain, the doped compound shows a drastic suppression of spectral weight  below $\hbar \omega\sim 8$~meV. The pseudogap is actually already visible in the raw TOF data Fig.~\ref{fig:data}(a). By analogy with Ca-doped \SCO \cite{Hammerath2011}, we initially suspected that the pseudogap may be a result of a spin-Peierls-like phase transition. However, no sign of any structural transitions were found in either calorimetric or Raman spectroscopy measurements\footnote{The Raman spectra were taken on the TriVista spectrometer from Princeton Instruments using $b(cc)-b$ geometry in the range 6 - 300 K and found to be essentially the same as in the parent compound \protect\cite{Misochko1996}.}.

Instead, we realized that the pseudogap can be explained as a generic feature of a single spin chain with randomly distributed defects. The behavior of a single defect has been considered in detail by Eggert and Affleck \cite{Eggert1992}. In particular, they showed that a chain with a $S=1$ impurity in place of an original $S=1/2$ spin is renormalized, in the limit of infinite chain length, to a {\it broken} $S=1/2$ chain,  and a totally {\it screened} (magnetically inactive) $S=1$ spin. The remaining low-energy physics is explained in terms of  finite-length chain segments. The magnetic susceptibility, for instance,  is primarily due to the paramagnetic contribution of odd-length segments \cite{Sirker2007,Sirker2008,Sirker2009}.
Indeed, $\chi(T)$ measured in our \SCON samples,  can be very well described by Curie behavior of free $S=1/2$ spins at a concentration $p=0.0037(1)$. This value is only slightly below the expected percentage of odd-length segments, $p=x/2=0.005$. The discrepancy can be attributed to deviations of defect concentration from nominal, as in the case of Sr$_2$Cu$_{1-x}$Pd$_x$O$_3$ \cite{Kojima2004,Sirker2007}, or to a weak temperature dependence of the effective Curie constant of fragmented chains \cite{Sirker2008}.
Another possibility is next nearest interactions in the chains that would tend to ``patch up'' the defects. Note that the double-chain structure of \SCO is an unlikely culprit, as it will have little impact on $\chi(T)$ for $T\gg J'x^2/k_\mathrm{B}\sim 0.01$~K \cite{Sirker2008}, which is clearly our case.

The main message of this work is that the spin dynamics can also be fully explained by the effective fragmentation of the spin chains by the screened $S=1$ defects. For a segment of length $L$, its ground state is separated from the first excited state by an energy {\it gap} $\Delta_L=\Delta_0/L$ for $L\gg 1$, with $\Delta_0\sim3.65J$ \cite{Eggert1992}. The defects are randomly positioned, which results in a distribution of chain segments lengths, and therefore a distribution of gap energies. In a macroscopic sample, this amounts to a pseudogap.

To quantify this effect, we need to calculate the momentum-integrated structure factor $\mathcal{S}_L(\omega)$ for a segment of arbitrary length $L$. This is a non-trivial task in the general case, but the basic physics can be captured by a rather simple argument. The elementary excitations in a $S=1/2$ Heisenberg AF chain are spinons \cite{Faddeev1981}, that have a linear dispersion at low energies. This part of the neutron spectrum is almost entirely due to a 2-spinon continuum \cite{Bougourzi1996}. On a finite-length segment the quasi-momentum of the spinons will be quantized: $q_{\|,n}=\pi n/(L+1)$, $n=1...L$. This will lead to a discrete excitation spectrum. Since the low-energy spinon dispersion is linear, the allowed energy levels will be equidistant: $\hbar \omega_L^n= n\Delta_L$. Thus, at low energies, the momentum-integrated spectrum can be represented as a series of regularly spaced $\delta$-functions:
\begin{equation}
\mathcal{S}_L(\omega)= \sum_n\mathcal{S}_L^n\Delta_L\delta(\hbar\omega-\hbar\omega_L^n),\label{e1}
\end{equation}
It stands to reason that for large $L$,
\begin{equation}
\mathcal{S}_L^n\approx \mathcal{S}_\infty(\omega_L^n),\label{e2}
\end{equation}
where $ \mathcal{S}_\infty(\omega)$ is the local structure factor for an infinite-length chain, Eq.~(\ref{sqw}). For a totally random distribution of defects with a concentration $x$, the probability to find a non-interrupted segment of length $L$ is given by $P_L=x^2(1-x)^L$. Summing up contributions of the distribution of length-$L$ segments from Eqs. (\ref{e1}) and (\ref{e2}), and by taking a continuum limit in $L\sim 1/x\gg1$, an approximation well-justified for $\hbar \omega \ll J$, allows us to arrive
at a fully analytical expression of the momentum-integrated structure factor of the diluted chain:
\begin{eqnarray}
\mathcal{S}(\omega) & \approx & \mathcal{S}_\infty(\omega)\times \sum_{n=1}^\infty n \left( \frac{\Delta}{\hbar\omega}\right)^2\exp\left(-n\frac{\Delta}{\hbar\omega}\right)\nonumber\\
&=& \mathcal{S}_\infty(\omega)\times F_\Delta(\omega)
,\label{SW}
\end{eqnarray}
where we have introduced the ``typical'' gap $\Delta=x\Delta_0$ and the defect-induced pseudogap ``envelope'' function
\begin{equation} F_\Delta(\omega) =\left(\frac{\Delta}{2\hbar\omega}\right)^2\sinh^{-2}\left(\frac{\Delta}{2\hbar\omega}\right).\label{env}
\end{equation}

This simple model has {\it no adjustable parameters}, yet is able to qualitatively reproduce the experimental data on \SCON. Expression (\ref{SW}) was evaluated for the nominal Ni-concentration of  $x=0.01$, assuming $\Delta_0=3.65J=0.825$~eV. The result is shown by the heavy solid curve in Fig.~\ref{fig:models} for a direct comparison with experiment. We most certainly do not expect a perfect agreement, in part due to our empirical treatment, but also due to an incomplete screening on the Ni-spins when they are close, a possible deviation of the actual Ni content in crystalline samples from the nominal value in the starting material, or to weak next-nearest-neighbor interactions. For the particular case of \SCON though, a much greater source of uncertainly is its double-chain structure. Inter-chain coupling, though frustrated, implies that each Ni-defect may act as a scattering center for low-energy spinons in {\it two} adjacent chains. In this case, the {\it effective} defect concentration $x$ should be doubled. In fact, assuming $x=0.02$ in our model gives a better quantitative agreement with experiment. To empirically account for all these effects, we chose to use the effective concentration as an adjustable parameter. The best fit to our low-temperature data corresponds to $x=0.017(1)$ and is shown by a dashed line in Fig.~\ref{fig:models}. This effective value will be used throughout the discussion below.

\begin{figure}
\includegraphics[width=\columnwidth]{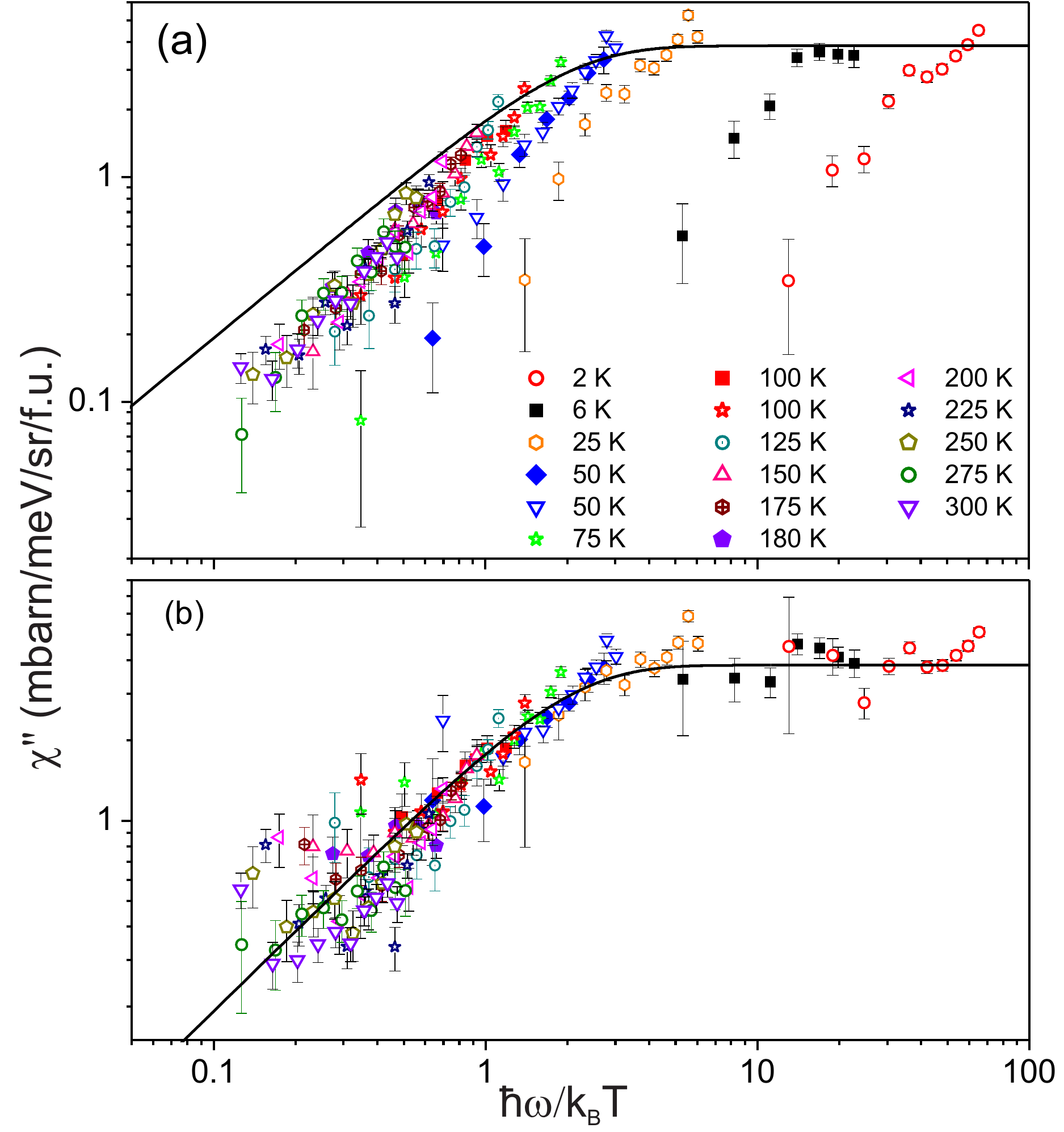}
\caption{(Color online)  (a) The imaginary part of generalized susceptibility measured at several temperatures (symbols). The plot is in coordinates where $\chi''$ of a simple spin chain is a universal scaling form given by Eq.~(\ref{sqw})(solid line) after correcting by population factor. (b) Collapse obtained for the same data, but normalized by the envelope function given by Eq.~(\ref{env}) with $x=0.017$. Open and solid symbols correspond to 3-axis and TOF data, respectively.\label{fig:scaling}}
\end{figure}

 A very intersecting feature of the model's prediction, one which is also consistent with experiment, is the extreme depletion of spectral weight at low energies, below about 2~meV transfer. This behavior is due to the exponentially low probability of having very long uninterrupted chain segments in the diluted system.

Perhaps the most convincing argument in favor of our interpretation, is that the {\it same} envelope function in Eq.~(\ref{env}), combined with Eq.~(\ref{sqw}), can {\it simultaneously} reproduce the measurements at all temperatures between 1.5~K and 300K. Specifically, the solid lines in Fig.~\ref{fig:data}b are obtained from Eq.~(\ref{SW}) assuming $x=0.017$ without any adjustable parameters, and are in good correspondence with the experimental data. The universal nature of the envelope function can be further demonstrated by considering that Eq.~(\ref{sqw}) has a scaling form, in that it depends only on $\hbar\omega/\kappa_\mathrm{B}T$. For the defect-free chain this is a consequence of a general finite-$T$ scaling for the Tomonaga-Luttinger spin liquid \cite{Giamarchibook}, and has been tested in a number of materials \cite{Dender1997thesis,Lake2005,Zheludev2007-2}. As can be seen in Fig.~\ref{fig:scaling}a, in our case of \SCON, the $\hbar\omega/\kappa_\mathrm{B}T$  scaling totally breaks down. This is due to the emergence of an additional energy scale, namely the pseudogap. However, if the data are normalized by the pseudogap envelope function (\ref{env}) , the scaling is {\it restored in the entire studied temperature range}, Fig.~\ref{fig:scaling}b.

Our approach is quite generic, and can be applied to describing the spin dynamics in chains with other types of defects, such as spin-vacancies or broken bonds. This brings us back to our initial question of how the pseudogap behavior can be reconciled with the divergent low-energy density of states in the supposedly universal RS phase? The key is that the strong disorder RG decimation procedure relies on always selecting the strongest AF bond in the chain. It can not be applied if all bonds are either severed, or equal to $J$. The diluted spin chain therefore does not flow to the RS state. As for the low-energy degrees of freedom that we expect to be released by disorder, in our case they are the free spins associated with odd-length segments, and take no part in the spin dynamics. The breakdown of the RS picture will have consequences for many real disordered spin chain materials, and may also account for the unexplained behavior of $S=1/2$ chains with weak-bond defects, such as in Cu(py)$_2$(Br$_{1-y}$Cl$_{y}$)$_2$
\cite{Thede2012}.

In summary, we have shown that Ni-doping of \SCO opens a pseudogap in the spin excitation spectrum. It can be generically explained in terms of site defects in the one-dimensional Heisenberg model, yieldind a universal scaling behavior, and without the involvement any structural phase transitions. In the future, it will be revealing to perform similar inelastic neutron scattering studies on \SCOC, as well as on doped versions of the single-chain compound Sr$_2$CuO$_3$.

Work at ETHZ was partially supported by the Swiss National Fund through MANEP. The work at IFW has been supported by the Deutsche Forschungsgemeinschaft (FOR 912) and the European Commission through the LOTHERM project (Project No. PITN-GA-2009-238475). Research conducted at ORNL's Spallation Neutron Source was sponsored by the Scientific User Facilities Division, Office of Basic Energy Sciences, US Department of Energy. The work of A. L. C. was supported by the US DOE under Grant No. DE-FG02-04ER46174, IZ by the US DOE Office of Basic Energy Sciences under Contract DE-AC02-98CH10886. We acknowledge support from the European Union under the Integrated Infrastructure Initiative for Neuron Scattering and Muon Spectroscopy (NMI3). AZ would like to thank D. Ivanov (ETHZ) for enlightening discussions. CH and BB thank H. Grafe, S. Nishimoto and S.L- Drechsler for discussions.


\end{document}